\def\YBCO{YBa$_2$Cu$_3$O$_{7-\delta}$}
\def\YBCs{YBa$_2$Cu$_3$O$_7$}
\documentstyle[prb,aps,multicol]{revtex}
  \renewcommand{\narrowtext}{\begin{multicols}{2} \global\columnwidth20.5pc}
  \renewcommand{\widetext}{\end{multicols} \global\columnwidth42.5pc}
  \newcommand{\wide}{\widetext \noindent \line(200,0){245} \line(0,1){3}\\}
  \newcommand{\narrow}{\begin{flushright}\mbox{\line(0,-1){3}$\! \!$
        \line(1,0){245}} \end{flushright} \narrowtext \noindent}

\input{psfig}
\draft{}
\begin{document}

\title
{Spatially-resolved studies of chemical composition,\\
critical temperature, and critical current density of\\
a YBa$_2$Cu$_3$O$_{7-\delta}$ thin film
}

\author{ M. E. Gaevski, A. V. Bobyl, D. V. Shantsev, R. A. Suris,
        V. V. Tret'yakov}
\address{
        A. F. Ioffe Physico-Technical Institute, Polytechnicheskaya 26,
        St.Petersburg 194021, Russia}
\author{ Y. M. Galperin,\thanks{Also with A. F. Ioffe
        Physico-Technical Institute} and
        T. H. Johansen,\thanks{E-mail: t.h.johansen@fys.uio.no}  }

\address{Department of Physics, University of Oslo, P. O. Box 1048
        Blindern, 0316 Oslo, Norway}
\maketitle

\begin{abstract}

Spatially-resolved studies of a YBa$_2$Cu$_3$O$_{7-\delta}$ thin film bridge
using electron probe microanalysis (EPMA), low-temperature scanning
electron microscopy (LTSEM), and magneto-optical flux visualization (MO)
have been carried out.
Variations in chemical composition along the bridge were measured by
EPMA with 3~$\mu$m resolution.
Using LTSEM the spatial distributions of the critical
temperature, $T_c$, and of the local transition width, $\Delta T_c$,
were determined with 5~$\mu$m resolution. Distributions of magnetic
flux over the bridge in an applied magnetic field have been measured
at 15 and 50~K by magneto-optical technique. The critical current density
$j_c$ as a function of coordinate along the bridge was extracted from
the measured distributions by a new specially developed method.
 Significant
correlations between $j_c$, $T_c$, $\Delta T_c$ and cation composition
have been revealed. It is shown that in low magnetic fields
deviation from the stoichiometric composition leads to a decrease in
both $T_c$ and $j_c$. The profile of $j_c$ follows the $T_c$-profile on
large length scales and has an additional fine structure on short scales. The
profile of $j_c$ along the bridge normalized to its value at any point
is almost independent of temperature.

\end{abstract}


\narrowtext
\section{Introduction}

The complicated crystal structure of high-$T_c$ superconductors (HTSC)
leads to their substantial spatial inhomogeneity which is specially
important because of the very short coherence length in those
materials. Consequently,
spatially-resolved studies of HTSC are very effective both to
evaluate the general quality of the samples and to determine local values of
important parameters. The quantities measured in the experiments which
do not allow spatial resolution are averaged over rather broad
distributions. Moreover, in some cases the properties of the whole 
sample can  be determined by one or few ``bottlenecks''. This appears to 
be one of the main obstacles to  adequate interpretation of 
experimental data and optimization of the performance of 
superconductor devices.    
  
Only a combination of different spatially-resolved methods allows one 
to relate different physical properties of the material in order to
facilitate the  
development of reliable theoretical models. As examples of such 
combinations several works can be mentioned. In 
Ref.~\onlinecite{bobylJAP}, spatially-resolved X-ray analysis together 
with measurements of voltage flicker noise allowed the study of the
relation between the noise 
level and a distribution of microstrains, in order to work out a relevant
theoretical model. Analysis of the correlation between locally 
measured cation composition, critical temperature, and flicker noise 
allowed the development of a theoretical model for cation defect formation 
in YBa$_2$Cu$_3$O$_7$ films\cite{bobylPhC}.     

Spatial distribution of the critical current
density, $j_c$, is also of great interest. It is a quantity that is
important for both HTSC applications and understanding the pinning
mechanisms. Distribution of $j_c$ can be inferred from the
distributions of magnetic field measured,
e.g. by magneto-optical (MO) imaging.
Unfortunately, most MO studies are restricted to a qualitative analysis
of magnetic field distributions since they are quite complicated
even for a homogeneous superconductor (see
Ref.~\onlinecite{mo} and references therein for a review).
Only few works\cite{mo1,LarbPhC} have been devoted to
analyzing the distributions of current density, $j$, restored
from MO images. The results give evidence of an extremely inhomogeneous
$j$-distributions and facilitate revealing factors limiting
current density. Extensive efforts in this direction seem to be crucially
important for subsequent progress in creation of
high-$j_c$ HTSC structures.\cite{LarbReview}

In this work we present a quantitative study of $j_c$-inhomogeneity
along a HTSC bridge using the MO technique. By means of low temperature
scanning electron microscopy (LTSEM), the spatial distribution of the
critical temperature, $T_c$, has been measured for the same
bridge. Simultaneous use of MO and LTSEM has earlier proved to be
successful for predicting the locations in a thin film bridge where
burn-out is caused by a large transport current\cite{burning}. The
present paper   reports the results of a comprehensive quantitative
investigation  of the correlation between the spatial distributions of
$j_c$, $T_c$ and chemical composition.

\section{Experimental}  
  
\subsection{Sample preparation}  
  
Films of YBa$_2$Cu$_3$O$_{7-\delta}$ were grown by dc magnetron sputtering
\cite{kar1} on LaAlO$_3$ substrate. X-ray analysis and Raman  
spectroscopy confirmed that the films were   
$c$-axis oriented and had a high structural perfection. Several  
samples, shaped as a bridge,   
were formed by  a standard lithography procedure. One of them, with
dimensions $460\times 110\times 0.2$ $\mu$m$^3$,   
was used for the present studies.  The absence of pronounced weak
links, and other defects which reduce the total   
critical current $I_c$, was confirmed by means of LTSEM\cite{LTSEM},  
and MO imaging.   
 The critical current density, $j_c$, determined   
by transport measurements was larger than  $10^5$~A/cm$^2$ at 77~K.  
The critical temperature defined by the peak of the
temperature derivative of resistance $dR/dT$ was $T_c=92.2$~K.  
The transition width defined by the width of $dR/dT$ peak  
was $\Delta T_c=2.2$~K.   
  
\subsection{Quantitative electron probe microanalysis}  
  
Spatially-resolved measurements of chemical composition and film
thickness have been performed using electron probe microanalysis on  
an X-ray spectral microanalyzer Camebax\cite{bobylPhC}. The electron  
energy in the exciting beam was 15~keV which allowed us to register  
simultaneously the spectral lines Y\ L$\alpha$, Ba\ L$\alpha$,  
Cu\ K$\alpha$, and O\ K$\alpha$ for all the elements,
and determine the film thickness in the same experiment. The absolute 
accuracy of the chemical composition   
determination is 0.3, 1.0, 1.2, and 2.0\% for    
Y, Ba, Cu and O, respectively. Such accuracy has been achieved by a  
special computer program to calculate the distribution of X-ray  
emission from both   
thin film and substrate under irradiation by the electron beam (see 
Ref.~\onlinecite{bobylPhC} for details).
The spatially-varying part of the composition is determined with three
times better accuracy for all the
elements due to improving  the measurement
procedure as follows: (i) we use computerized control of the probe
current while  accumulating
10$^5$ pulses from the X-ray detector, (ii) a special
computer  program enabling composition determination at 150 points of
the film during
one experimental run was implemented, and (iii) we performed a running
calibration based on
comparison of the line intensities with those of  a pure \YBCO\ single
crystal placed in the same chamber. The spatial resolution of the
method is 3~$\mu$m.

\subsection{Low-temperature scanning electron microscopy}

The LTSEM technique originally
developed to study conventional
superconductors\cite{ClemHuebener,LTSEM} has recently been adapted for
determination
of spatial distributions of critical temperature, or
$T_c$-maps\cite{water,izvan}.
The method is based on monitoring the local transition into the normal
state due to heating by a focused electron beam.
Heating by the beam elevates the
temperature locally causing a change in
the local resistivity. As a result,  a change
in the voltage occurs across the sample biased by a transport
current. Since the  electron beam
induced voltage (EBIV) is proportional to the temperature derivative of
the local resistivity, the signal reaches its maximum at the
temperature equal to the local value of $T_c$. The width of the
maximum corresponds to the local transition width, $\Delta T_c$.
Scanning the electron beam over the film allows us to  
determine the spatial distribution of both $T_c$ and $\Delta T_c$. The 
spatial resolution for $T_c$ and $\Delta T_c$ determination can be 
improved by a proper treatment   
of the EBIV distributions taking into account  heat diffusion from the  
irradiated region into surrounding areas.  
 As a result, a spatial resolution up to 2~$\mu$m and a temperature
resolution up to 0.2 K can  be achieved\cite{water,izvan}.

The LTSEM measurements were carried out with an automated scanning  
electron microscope CamScan Series 4-88 DV100 equipped with a   
cooling system ITC4 and a low-noise amplifier for voltage  
signals. The temperature could be maintained to within
0.1~K in the range 77-300~K. The bias current $I$ was varied from 0.2 
to 2.0~mA so that the value of $I$ was large enough to detect EBIV and 
small enough to avoid distortion of the transition  by bias 
current. EBIV was measured using a simple four-probe scheme.    
  To extract the local EBIV signal, lock-in detection was used with a
beam-modulation frequency of 1 kHz. The electron beam current was
10$^{-8}$~A while the acceleration voltage was 10~kV.  
  
\subsection{Magneto-optical technique}  
  
Our system for flux visualization is based on the Faraday rotation of  
a polarized light beam illuminating an MO-active indicator film  
placed directly on top of the sample  surface. The rotation angle
grows with the magnitude of the local magnetic field perpendicular to  
the HTSC film, and by using crossed polarizers in an optical  
microscope one can directly visualize and quantify the field  
distribution across the sample area. As Faraday-active indicator we  
use a  Bi-doped yttrium iron garnet film with in-plane  
anisotropy\cite{dor1}. The indicator film was deposited to a thickness  
of 5 $\mu$m by liquid phase epitaxy on a gadolinium gallium garnet
substrate. Finally, a thin layer of aluminum was evaporated onto the  
film in order to reflect the incident light and thus providing a  
double Faraday rotation of the light beam.   
 The images were recorded with an eight-bit Kodak DCS 420 CCD camera  
and transferred to a computer for processing.
 After each series of measurements at a given temperature, the
temperature was increased above $T_c$ and an in-situ calibration of  
the indicator film was carried out.  
 As a result, possible errors caused by inhomogeneities of both  
indicator film and light intensity were excluded. The experimental 
procedure is described in more detail in Ref.~\onlinecite{joh1}.

\section{Results}  

To report the experimental results we employ the following notations.
The $x$-axis is directed across the bridge, the edges being located at
$x=\pm w$, the $y$-axis points along the bridge,
and the $z$-axis is normal to the film plane.
In what follows, the distributions of the chemical compositions,
$T_c$, $\Delta T_c$
and $j_c$ are analyzed as functions of $y$.

\begin{figure}
\centerline{\psfig{figure=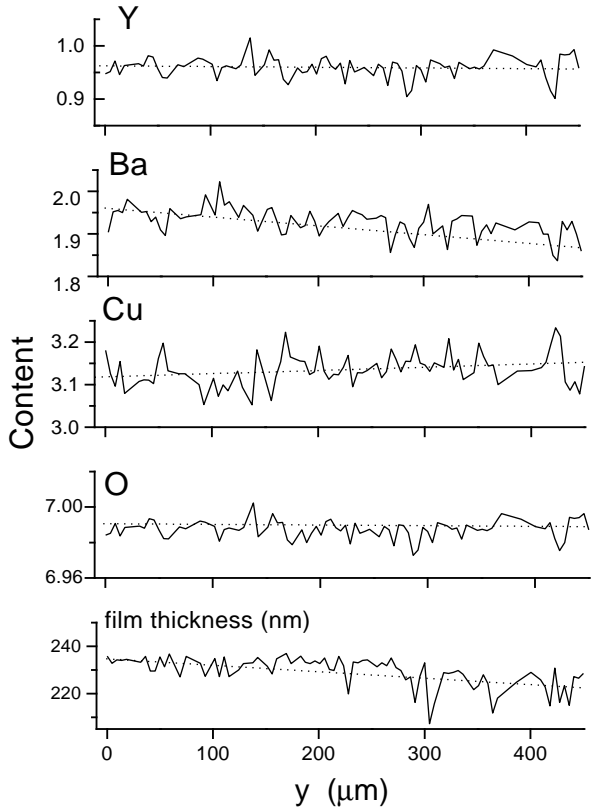,width=8.2cm}}
\caption{Variations in microscopic chemical composition along \YBCO\
bridge and variations in film thickness measured by EPMA. Dotted lines
show linear fits to data. A gradient in Ba and Cu content is clearly
seen.}
\label{f_cation}
\end{figure}

\subsection{Chemical composition variation}

\begin{figure}
\centerline{\psfig{figure=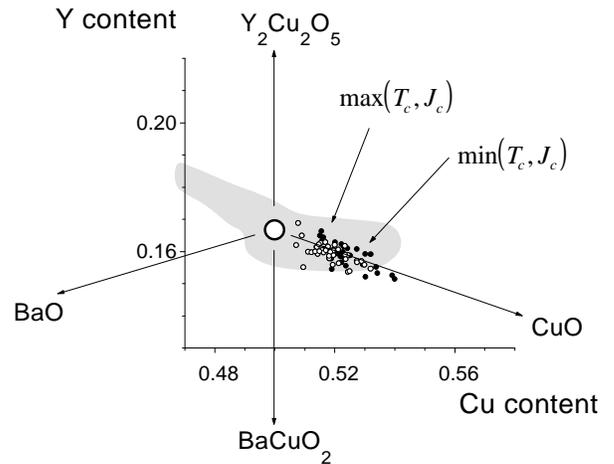,width=8.2cm}}
\caption{Composition diagram of Y$_y$Ba$_{1-x-y}$Cu$_x$O$_z$  in the
vicinity of
stoichiometric \YBCs\ composition projected on the plane $z=0$. The
stoichiometric composition is shown by the large open circle. Small
open (solid) circles correspond to local compositions measured in the
left (right) part of the bridge. The lines show
directions towards known stable compounds. Gray area
shows the region{\protect{\cite{bobylPhC}}} where superconducting
properties of the material are only weakly sensitive to the
composition. The closer local composition to the stoichiometric one,
the higher local values of $T_c$ and $J_c$.}
\label{f_phased}
\end{figure}

Variations in Y, Ba, Cu, and O contents along the \YBCO\ bridge are shown
in Fig.~\ref{f_cation}.  A systematic
gradients in Ba and Cu content are clearly visible and
indicated by the dotted lines representing linear fits to the data.
Note, that the left part of the bridge (small $y$)
is closer to the
stoichiometric composition, \YBCs, than the right part.
It can also be seen from Fig.~\ref{f_cation} that,
in contrast to the
cations,
the oxygen is  distributed  rather uniformly over the bridge. A
uniform oxygen distribution in \YBCO\ films has been observed
also earlier\cite{bobylJAP,bobylPhC}. It is probably a consequence of high
diffusion coefficient of oxygen in the  \YBCO\ lattice.  Thus, we can focus
on variations in the cation composition only. The composition diagram
for   Y$_y$Ba$_{1-x-y}$Cu$_x$O$_z$  in the
vicinity of  stoichiometric \YBCs\ composition projected on the plane
$z=0$ is  shown in Fig.~\ref{f_phased}.
The open (solid) data points correspond to local compositions measured
in the left (right) part of the bridge.

Except the long-scale gradient in Ba and Cu, short-scale oscillations
in the cation composition can be seen from Fig.~\ref{f_cation}. A
careful analysis of the data shows that short-scale oscillations in Y
and Ba content are correlated with each other and anti-correlated with
those in Cu content. As a result, the experimental data points in
Fig.~\ref{f_phased} are mainly spread along the direction towards CuO
oxide. To clarify the origin of this phenomenon
\noindent
we carried out SEM
studies of the bridge surface which revealed the presence of submicron
CuO inclusions. Inclusions are formed due to excess of Cu in \YBCs\
lattice and lead to a slightly nonuniform distribution of Cu on
micron scale.  Note that these short-scale variations in cation
composition have nothing to do with the long-scale gradient in Ba and Cu
content. In this work we are interested in the long-scale composition
variations only. Below they will be compared to the long-scale variations
in $T_c$ and $j_c$.

\subsection{$T_c$ and $\Delta T_c$-profiles}

\begin{figure}
\centerline{\psfig{figure=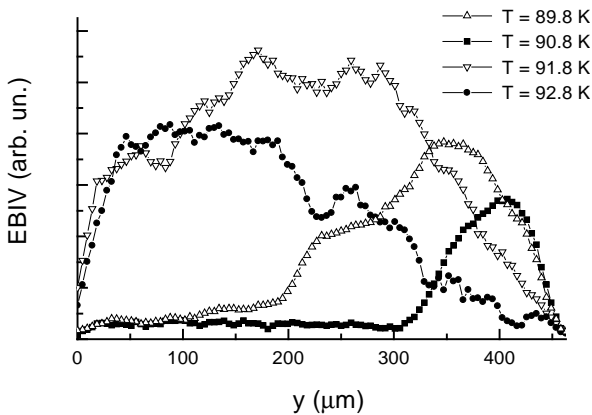,width=8.5cm}}
\caption{Profiles of the EBIV measured by LTSEM along the \YBCO\
bridge for different temperatures. Essential spatial inhomogeneity of
EBIV is clearly seen.}
\label{f_ebiv4t}
\end{figure}

The EBIV profiles along the bridge for four temperatures
 are presented  in Fig.~\ref{f_ebiv4t}.
Each point is obtained by averaging local
EBIV along the $x$-direction, i.e., over the bridge cross-section.
A systematic inhomogeneity of the bridge can be seen.
 Large EBIV for higher temperatures in the left part of
the bridge corresponds to  higher $T_c$ there,
while in the right part, EBIV is large
at low temperatures indicating lower $T_c$.

\begin{figure}
\centerline{\psfig{figure=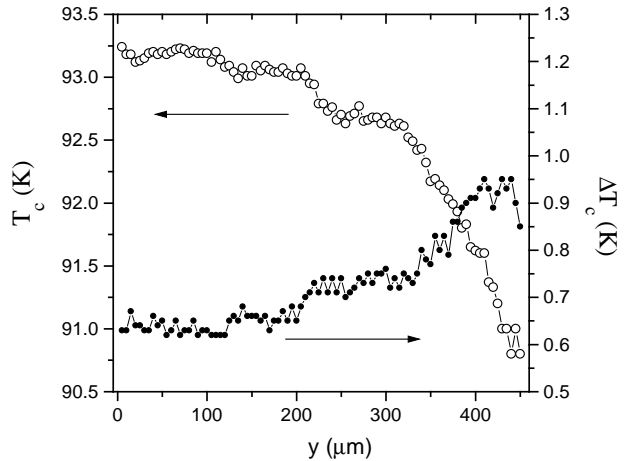,width=8.5cm}}
\caption{Profiles of the critical temperature, $T_c$, and of the
transition width, $\Delta T_c$, along the bridge determined from LTSEM
data.}
\label{f_tc(y)}
\end{figure}

The spatial distributions of the critical temperature $T_c$
and the transition width, $\Delta T_c$, with 5~$\mu$m resolution
has been determined according to the
procedure mentioned in Sec.~II\ C and described in more detail in
Ref.\onlinecite{water} and~\onlinecite{izvan}.
The profiles $T_c(y)$ and $\Delta T_c(y)$, as shown in Fig.~\ref{f_tc(y)},
have been calculated by averaging over 20 points across
the bridge.
The standard deviation is less than the actual accuracy of the $T_c$ and
$\Delta T_c$ determinations which is $0.2$~K.
One can clearly see a gradient in $T_c$ which is especially large in
the right part of the bridge. Note that a decrease in $T_c$ is
accompanied by an
increase in $\Delta T_c$. Larger transition width, $\Delta T_c$, in
the right part of the bridge is most probably related to an
inhomogeneous distribution of $T_c$ on the scales shorter than LTSEM
resolution, 5~$\mu$m. Such a short-scale $T_c$-inhomogeneity can
hardly be expected in the left part of the bridge where $T_c$
approaches its maximal value,  $\ge 93$~K, corresponding
to a
stoichiometric composition of the material.

It should be noted that only the values of $T_c$ greater than some
minimum temperature, $T_{\min} \approx 90.7$ K, can be determined by
the present method. We believe that the  temperature
$T_{\min}$ corresponds to formation of superconducting percolation
cluster, and at  $T<T_{\min}$ the EBIV falls below
our experimental resolution.
The results presented in Fig.~\ref{f_tc(y)} are obtained
by averaging over the regions with $T_c > T_{\min}$ corresponding to
66\% of the area for our sample.
The value $p_c=66\%$ for the percolation threshold seems reasonable since
for an infinite random two-dimensional system $p_c \approx 50\%$ and
the bridge shape seems intermediate between a 2D and a 1D geometry.

\subsection{MO results: $j_c$-profiles}

\begin{figure}
\centerline{\psfig{figure=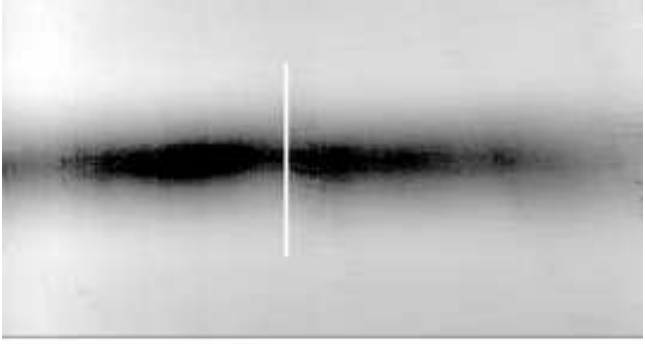,width=8.5cm}}
\caption{Magneto-optical image of flux distribution in a bridge in a
perpendicular
external field of $B_a=21$~mT at temperature $T=15$~K. Dark regions
correspond to low flux density. In the right part of the bridge
the width of the dark region is reduced indicating deeper flux penetration
and, hence, lower critical current
density. A region of deep flux penetration in the center of the
bridge is marked by a white line. }
\label{f_image}
\end{figure}

\begin{figure}
\centerline{\psfig{figure=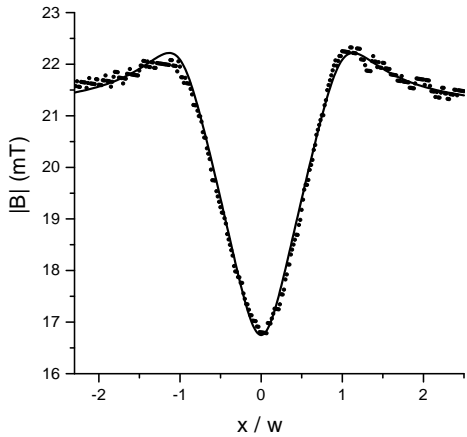,width=7cm}}
\caption{A typical magnetic field profile across the bridge
averaged over distance $2w$ along the bridge.
The external field was $B_a=21$~mT. The line shows a fit by the
Bean model, Eq.~(\ref{Bx2}) using $j_c=4.7\cdot 10^{6}$~A/cm$^2$,
and $h=0.33 w$.}
\label{f_bprof}
\end{figure}

The magnetic field distributions in perpendicular
applied fields
up to 35~mT have been measured using the MO  technique
at $T=15$~K and 50~K. A typical MO image of the narrow strip part of
the bridge is shown in Fig.~\ref{f_image}.
Figure~\ref{f_bprof} shows a typical profile of the
absolute value of the $z$-component of magnetic induction
across the bridge for an external field $B_a = 21$~mT.
The profile is obtained by averaging the flux
distribution over a 110~$\mu$m length along the bridge.

The data were fitted to the Bean model for
thin strip geometry\cite{BrIn,zeld1}. For the indicator film placed at
the height $h$ above the bridge the $z$-component of magnetic
induction is given by the expression~\cite{joh1},
\widetext
\begin{eqnarray}
  B(x) &=& \frac{B_c}{4}
\left\{ \ln \frac {
                    \left[ (x+a)^2 + h^2 \right]
                    \left[ (x-a)^2 + h^2 \right]
                   } {
                    \left[ (x+w)^2 + h^2 \right]
                    \left[ (x-w)^2 + h^2 \right]}
%\right.\nonumber \\ &&\left.
- \frac 4\pi \int^a_{-a}
\frac{x'-x}{(x'-x)^2+h^2}
\arctan \left(
\frac {x'}w \sqrt{\frac{w^2-a^2}{a^2-x'^2}}
\right) dx' \right\} + B_a.
\label{Bx2}
\end{eqnarray}
\narrow
Here
\begin{equation}
a=w/\cosh \left(B_a/B_c \right), \quad
 \quad B_c \equiv \frac{\mu_0 J_c}{\pi}\, .
\label{a}
\end{equation}
The quantity $a$ limits the area
of field penetration (the region $|x|<a$ is vortex free). The sheet 
critical current density, $J_c$ is defined as $J_c=\int^d_0 j_c(z)\, 
dz$, where $d$ is the film thickness.   
  
The contact pads which are necessary for the LTSEM  
measurements screen the applied field to some extent. As a result,  
the actual external field $B_a$ acting upon the bridge is unknown. Therefore,
Eq.~(\ref{Bx2}) contains three unknown quantities, $B_c$, $h$, and  
$B_a$. Fortunately, the situation appears rather simple when $B_a \gg  
B_c$, as $a$ then becomes negligible.   
For $B_a > 3 B_c$  substituting   
$a=0$ into Eq.~(\ref{Bx2}) leads to  $\le 1\%$ error  
in $B(x)$ for any $x$.  
 The quantity $B_a$ then enters Eq.~(\ref{Bx2}) as an  
additive constant, and we eliminate it by considering the difference  
$\delta B (x) \equiv B(x+\Delta) -B(x-\Delta)$. Here $\Delta$ is a constant  
shift which we chose to be equal $10 \ \mu$m. The expression for  
$\delta B(x)$ has the form $\delta B(x)=B_c {\cal L}(x,h)$ with
\wide
\begin{equation} \label{dBx}
{\cal L}(x,h) = \frac{1}{4} \, \ln \frac {
                    \left[ (x+\Delta)^2 + h^2 \right]^2
 \left[ (x-\Delta +w)^2 + h^2 \right]
                    \left[ (x-\Delta -w)^2 + h^2 \right]}
                    {\left[ (x-\Delta)^2 + h^2 \right]^2
                    \left[ (x+ \Delta+w)^2 + h^2 \right]
                    \left[ (x+\Delta-w)^2 + h^2 \right]
}
\end{equation}
\narrow
Thus, we are left with two unknown parameters, $B_c$ and $h$.
The experimental curves for $\delta B(x)$ were fitted by the formula
(\ref{dBx}), the parameters $B_c$ and $h$ being determined by
minimizing the quantity
$$S=\int_{-2w}^{2w} dx\, \left[ B_c {\cal L}(x,h)- \delta
B_{\text{exp}}(x)\right]^2\, .$$
The condition $\partial S/\partial B_c =0$ implies that the two
unknown parameters  are related by the expression
\begin{equation} \label{Bc}
B_c= \int_{-2w}^{2w} dx\,  \delta
B_{\text{exp}}(x) \, {\cal L}(x,h) \, \left[ \int_{-2w}^{2w} dx\,
{\cal L}^2(x,h) \right]^{-1}.
\end{equation}
{}From the fitting, the height $h$ was found to be a linear function of the
co-ordinate $y$,
\begin{equation} \label{h}
h =14\, \mu\text{m} +0.006\, y \, .
\end{equation}
This corresponds to a tilt angle of $\approx 0.3^\circ$ for the indicator
film with respect to the surface of the sample.

The fitting procedure identifies good agreement between experimental and
theoretical flux profiles. Figure~\ref{f_bprof} shows an example of
such a fit, which also verifies
the adequacy of the Bean model for our experimental situation.
Applying  higher external field we checked that this method, based on
assumption of a $B$-independent $J_c$, works
with a proper accuracy up to $B_a \approx 30$ mT. The method is
therefore applied to determine $J_c$ values in different
cross-sections of the  bridge.

It should be noted that the basic expression, Eq.~(\ref{Bx2}), is
derived  for a
homogeneous  infinite
strip\cite{zeld1,BrIn}. Consequently, it is valid only for a
smooth inhomogeneity along the bridge. One can expect that the
characteristic scale of inhomogeneities which can be analyzed using
Eq.~(\ref{dBx}) should be larger than the bridge width, $2w$.
To estimate
the accuracy of the employed method we have compared the magnetic
field $B(x,\infty)$ in an infinite
strip  and the field $B(x,L)$ in the middle of a
finite-length bridge $(-L \le y \le L)$. For the case of full penetration,
$J=J_c ( x/|x|)$, the difference between the fields at height
$h$  is given by the expression
\wide
\begin{equation} \label{dB}
  B(x,L)-B(x,\infty) = \frac{\mu_0 J_c}{4 \pi}
 \ln \frac { \left[\sqrt{(x+w)^2 + h^2 +L^2}+L \right]
 \left[\sqrt{(x-w)^2 + h^2 +L^2}+L \right]}
{(\sqrt{x^2 + h^2 +L^2}+L)^2}\, .
\end{equation}
\narrow
This expression reaches the maximum in the bridge center,
$x=0$. Substituting $h=0.33w$ and $L=w$ we find that $B(0,w) -
       B(0,\infty) \approx 0.15
B(0,\infty)$, i. e. the field for the strip with the length $2w$ is
$\approx 85\%$ of that for the infinite strip. Further, substitution
of $B(x,w)$ into Eqs.~(\ref{dBx}) and~(\ref{Bc}), leads to the
error about 9\% in the value of restored sheet current density
$J_c$. Thus, the proposed method allows determination of variations in
$J_c$ on length scales 2$w$ with accuracy better than 9\%.

      Based on the above estimates we have averaged the experimental
profiles $B(x)$ over the intervals $(y-w,y+w)$
\begin{figure}
\centerline{\psfig{figure=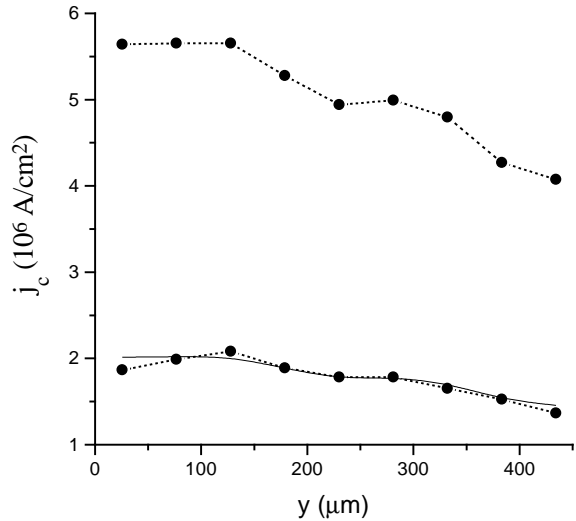,width=8.5cm}}
\caption{Profiles of the critical current density $j_c$ along the
bridge measured at 15~K (upper curve) and 50~K (lower curve). Values of $j_c$ were calculated  from the
measured magnetic field distribution
in external field $B_a=21$~mT. Solid line corresponds to $j_c(15
\,$K$)/2.8$.}
\label{f_jc(y)}
\end{figure}
for several $y$ and then
calculated the critical current density $j_c(y)=J_c(y)/d$ from
Eqs.~(\ref{Bc}) and~(\ref{a}). The results for  $T=15$ K
and  $T=50$ K are shown in
Fig.~\ref{f_jc(y)}. These results are obtained
from  $B(x)$-profiles
at  $B_a=21$~mT which we consider as an optimal value of external  
field. Indeed, at low applied fields $B_a$ our assumption $a=0$   
is not valid, while at high $B_a$, $j_c(B)$-dependence  
 becomes noticeable and the Bean model is not applicable. As seen from Fig.~\ref{f_jc(y)}, $j_c$ is essentially  
inhomogeneous. Values of $j_c$ at opposite edges  
of the bridge differ by almost a factor of two. Since the apparent value of  
the critical current density can be affected by variation of the film 
thickness $d$, we also measured the profile of $d$ along the bridge 
using the EPMA technique. It can be seen from the lower panel of   
Fig.~\ref{f_cation} that the variation in $d$ is about 5\%. Therefore,
it   cannot be responsible for the observed variation in $j_c$ since the
latter is substantially larger.
  
Note that the curves for $T=15$~K  
and  $T=50$~K  differ  
practically by a constant factor $\approx 2.8$. To illustrate this  
fact we show profile  of the critical current for 15 K divided by 2.8   
by solid line in Fig.~\ref{f_jc(y)}.

\section {Discussion}  
  
The main results of this work is the observation of a substantial 
sensitivity of both   
the  critical parameters of the superconductor, $T_c$ and $j_c$,  
to the material composition. As the composition deviates from the  
stoichiometric one towards excess of Cu and Ba deficiency, both $T_c$  
and $j_c$ decrease. Furthermore, although the composition varies  
gradually over the  
whole bridge, the critical parameters vary gradually in some region  
where the deviation from the stoichiometric composition is small, while
outside this region they decrease drastically (see
Figs.~\ref{f_cation},~\ref{f_tc(y)} and~\ref{f_jc(y)}).   
Such  behavior is consistent with existence of the region in   
vicinity of stoichiometric composition, \YBCs, where superconducting 
properties are only weakly sensitive to the composition. When   
composition falls beyond this region, the material contains defects  
which dramatically affect  the electronic  
 properties and in this way reduce significantly both $T_c$ and  
$j_c$. Some ideas regarding the shape of this region in  
\YBCO\ can be inferred from the studies of cation defect formation  
performed in Ref.~\onlinecite{bobylPhC}. An excess of Cu along with a  
Ba deficiency  leads to substitution defects which  
neither produce substantial strain in the lattice, nor cause charge  
redistribution. Therefore the width of the region is rather  
wide in the mentioned direction, that is illustrated by  
Fig.~\ref{f_phased}. Change of Ba content from 1.95 to  
1.85 is accompanied by only a 2~K decrease in $T_c$.  
\begin{figure}
\centerline{\psfig{figure=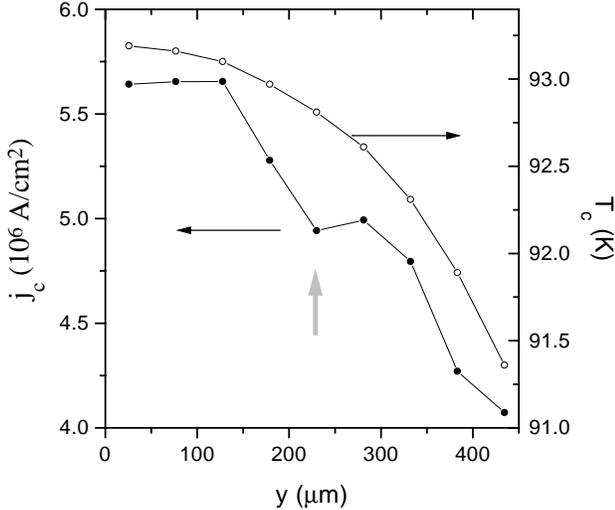,width=8.5cm}}
\caption{Profiles of the critical current density, $j_c$, at $T=15$ K,
and of the critical temperature, $T_c$, along the bridge. Arrow marks
the region of reduced $j_c$ which is also seen  in
Fig.~\ref{f_image} and marked by the white line. }
\label{f_jctc(y)}
\end{figure}

\begin{figure}
\centerline{\psfig{figure=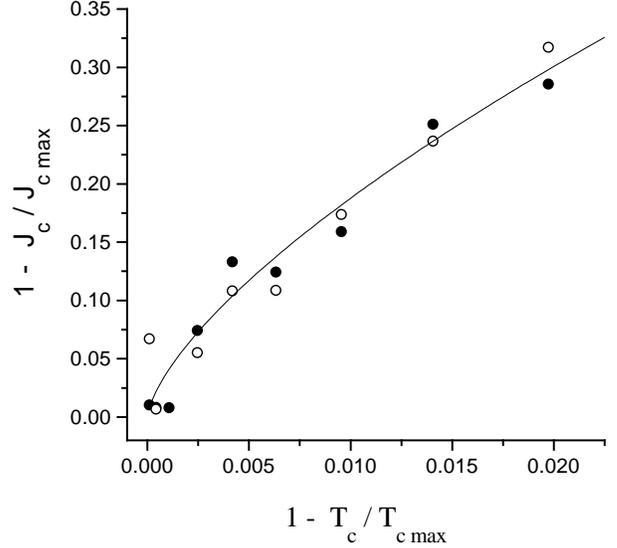,width=8.5cm}}
\caption{Correlation between the normalized critical current density,
$J_{c}$, and critical temperature. $J_{c\, \max}$ and $T_{c\, \max}$
are the maximal
values of $J_{c}$ and $T_{c}$ over the bridge. Open
circles correspond to $T=50$ K while the solid ones correspond to
$T=15$ K. Solid line shows the power law fit with the exponent $\nu=0.7$.}
\label{f_jctc}
\end{figure}

A clear correlation between $j_c$ and $T_c$ is illustrated by
Figs.~\ref{f_jctc(y)} and~\ref{f_jctc}. Despite of qualitative
similarity of the $j_c(y)$ and $T_c(y)$ dependences, the critical
current varies much more strongly; the variation in $T_c$ is $\approx 2\%$
while $j_c$ varies by almost a factor of two.   
To describe the correlation between $j_c$ and $T_c$ in a quantitative  
way let us note that $j_c (y)$ profiles for 15 and 50 K differ only by  
a numerical factor (see Fig.~\ref{f_jc(y)}). Thus the critical current  
at large enough scale can be described as $j_c (T,y)) =  
F(T)j_{c0}[T_c(y)]$, where $F(T)$ is a function of the temperature. It  
follows from Fig.~\ref{f_jc(y)} that $F(15 \,   
\text{K})/F(50 \, \text{K}) =  2.8$.   
In Fig.~\ref{f_jctc} the dimensionless quantity $(1-j_c/j_{c\,  
\max})$ is plotted versus the quantity $(1 -T_c/T_{c\,\max})$ for  
$T=15$ and 50 K. Here  
$j_{c\,\max}$ and $T_{c\,\max}$ are the maximum values  
of $j_c$ and $T_c$ over the bridge. The data can be approximated by a  
power law function with the exponent $\nu \approx 0.7$. Thus, one can  
express the $j_c-T_c$ correlation as                             
\begin{equation}  
\label{nu}
1-J_{c}(T)/J_{c\, \max}(T) \propto (1-T_c/T_{c\,
\max})^{\nu} \, ,
\end{equation}
with a temperature {\em independent} coefficient.
\begin{figure}
\centerline{\psfig{figure=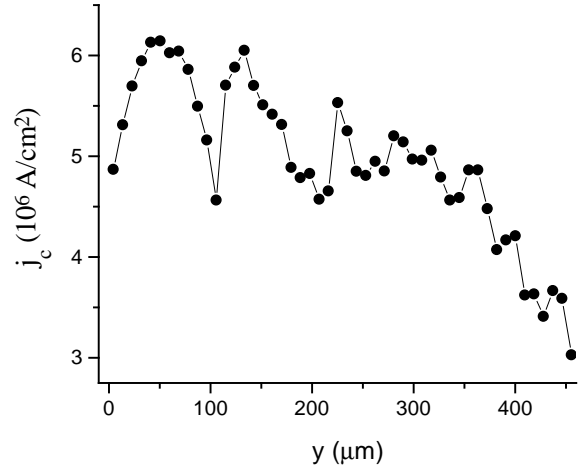,width=8.5cm}}
\caption{Profile of the critical current density $j_c$ along the
bridge. The method used for determination of $j_c$ provides accurate
results on length scales $\ge 110 \mu$m. Thus, the figure illustrates
the {\em existence} of $j_c$-inhomogeneity on short scales, but does
not give a {\em quantitative} information about this short-scale
inhomogeneity.}
\label{f_jcfine}
\end{figure}

Meanwhile, spatial dependence of $j_c$ possesses an additional fine structure
compared to the spatial dependence of $T_c$. In Fig.~\ref{f_jcfine}
the $y$-dependence of $j_c$ obtained from $B(x)$-profiles averaged
over 10 $\mu$m length along the bridge is
shown. This curve serves to demonstrate the character because the typical  
scale of $j_c$-inhomogeneity appears $\lesssim w$. On the other  
hand, according to the above estimates, our method for $j_c$ 
determination is quantitatively valid only   
for the scales $\ge 2w$. However, it is clear that there are rather   
pronounced inhomogeneity of $j_c$ at the scales $\lesssim 50 \ \mu$m.   
This inhomogeneity can be ascribed, depending on the mechanism of the
critical   
current, either to inhomogeneous pinning, or to inhomogeneity of weak  
links between superconducting regions.   
Meanwhile, the long-scale variation in $j_c$
correlated to the long-scale behavior of $T_c$ provides an evidence that
the critical current is substantially influenced by changes in
the electron properties caused by the deviation from stoichiometric
composition.   
  
There are numerous examples in the literature showing that  
introduction of  structural defects, e.g. by heavy ion irradiation,  
increases the critical current density.  
One could expect that deviation from stoichiometric composition would  
lead to formation of additional structural defects which may serve as  
pinning centers for magnetic flux and in this way increase $j_c$. On  
the other hand, structural defects lead to changes of electron properties  
which suppress superconductivity and decrease $j_c$. Results of this  
work suggest that composition-induced variation in  
electronic structure  influences the critical current density stronger than   
appearance of additional pinning centers caused by the deviation from the  
stoichiometry.   
This conclusion, however, is valid only for the applied range of 
magnetic fields, $B   
< 35$~mT. Indeed, as shown in Ref.~\onlinecite{gapud}, introduction of
structural defects may lead to decrease in the critical
current density $j_c$ at low magnetic fields and increase of $j_c$ at
high fields.
  
There are two main mechanisms limiting the current density in inhomogeneous  
superconductors -- intra-grain vortex depinning and suppression of  
Josephson effect in weak links between the grains.  
These mechanisms can be distinguished by analyzing the temperature  
dependence of the critical current\cite{jung}. Evidence that  
intra-grain $j_c$ has a stronger temperature dependence than  
inter-grain $j_c$ is provided by MO imaging of flux penetration into  
\YBCO\ crystal containing weak links at different  
temperatures\cite{welp}. The substantial decrease of $j_c$ with  
temperature observed in the present work supports intra-grain  
vortex depinning   
as the main mechanism limiting current density.   
This conclusion is  
in agreement with the results of Ref.~\onlinecite{polyanskii}.  
In that work the critical current density   
across a single grain boundary, $j_c^{\text{gb}}$, and the bulk
critical current density,
$j_c^{\text{bulk}}$, have been determined independently.
Use of a similar material (thin \YBCO\ films) and a similar method
of defining $j_c$, as well as comparable values of $j_c$, allows one to think  
that the results of  
 Ref.~\onlinecite{polyanskii} are relevant to our case.   
It has been shown\cite{polyanskii} that for a 7$^{\circ}$
misorientation angle boundary  
 $j_c^{\text{gb} }(15 K)/j_c^{\text{gb}}(50 K) = 1.4$, while for bulk  
critical current density:  
 $j_c^{\text{bulk}}(15 K)/j_c^{\text{bulk}}(50 K) = 2.3$.  
Our value $j_c(15 K)/j_c(50 K) = 2.8$ is closer to the case of bulk  
pinning which is probably the main mechanism   
limiting the current density in studied film.

It is worth to be emphasized that though there is a clear   
correlation  
between $j_c$ and $T_c$ the variation of $j_c$ with the composition is  
much stronger compared to the variation of $T_c$. This fact can
hardly  be understood from the BCS model which would predict
comparable relative variations of the above mentioned
quantities. Studies of electron band structure would probably
 give a key to understanding mechanisms responsible for observed $T_c$ 
and $j_c$ variations.

\section{Conclusion}  
  
Inhomogeneity of \YBCO\ thin film bridges is investigated using three 
experimental   
methods allowing spatially resolved measurements -- electron probe  
microanalysis, low-temperature scanning electron microscopy, and  
magneto-optical imaging. The profiles of chemical composition,  
critical temperature, and critical current density along the bridge are 
determined.   
  
It is shown  that in low magnetic fields, deviation from the  
stoichiometric   
composition leads to a decrease in both critical temperature and
critical current
density.
This fact allows to conclude that composition-induced variation in
electronic structure  influences the critical current density more
strongly than the   
appearance of additional pinning centers caused by the deviation from  
stoichiometry.    
Therefore, the way to optimize both parameters is to keep the  
composition as close to stoichiometric as possible.   
  
The profiles of the critical current density along the bridge at  
different temperatures appear to be proportional, i. e. they scale by 
a temperature-dependent factor. Consequently, the profile of critical 
current normalized to its value at any point is essentially 
independent of temperature.   
  
The profiles of the critical current density possess an additional  
fine structure at short scales ($<50 \, \mu$m) which is absent in profiles  
of $T_c$. This fine structure indicates that characteristic scale of 
the pinning strength
inhomogeneity is much less than that of the
$T_c$-inhomogeneity.

\acknowledgements  
  
The financial support from the Research Council of Norway and from the  
Russian National Program for Superconductivity is gratefully  
acknowledged.

\widetext
\end{document}